# Dynamic Clustering Protocol for Data Forwarding in Wireless Sensor Networks

Deepali Virmani, Akshay Jain , Ankit Khandelwal , Divik Gupta, Nitin Garg
Department of Computer Science & Engineering
Bhagwan Parshuram Institute of Technology
Delhi, India.
deepalivirmani@gmail.com

**Abstract-Energy being the very key concern area with sensor networks, so the main focus lies in developing a mechanism to increase the lifetime of a sensor network by energy balancing. To achieve energy balancing and maximizing network lifetime we use an idea of clustering and dividing the whole network into different clusters. In this paper we propose a dynamic cluster formation method where clusters are refreshed periodically based on residual energy, distance and cost. Refreshing clustering minimizes workload of any single node and in turn enhances the energy conservation. Sleep and wait methodology is applied to the proposed protocol to enhance the network lifetime by turning the nodes on and off according to their duties. The node that has some data to be transmitted is in on state and after forwarding its data to the cluster head it changes its state to off which saves the energy of entire network. Simulations have been done using MAT lab. Simulation results prove the betterment of our proposed method over the existing Leach protocol.**

## I .INTRODUCTION

Sensor networks are composed of small, battery operated sensors, whose main function is to collect and forward the required data to the base stations. WSN's facilitate monitoring and controlling of physical activities from the surveillance areas with better accuracy [1] [10]. WSN's have applications in a variety of fields such as environmental monitoring, military purposes and gathering sensing information in inhospitable locations [1] [10]. A sensor network is an infrastructure comprised of sensing (measuring), computing, and communication elements that gives an administrator the ability to instrument, observe, and react to events and phenomena in the specified environment. The administrator typically is a civil, governmental, commercial, or an industrial entity.

The WSN's are built of "nodes" that range from a few to several hundreds or even thousands, where each node is connected to one (or sometimes several) sensors. Every sensor has three basic units- sensing, radio, and battery, the major constraint being limited energy as the sensor nodes are directly dependent on the battery life [1] [11]. Till battery is charged they are live, so our focus lays on make them live for more amount of time. Consequently, a large number of sensor nodes can be networked to gather sensory data and each sensor performs two main responsibilities, namely, Sensing activities and Routing the sensed data to the base station or a controller (Gateway Sensor Node). The base station is a master node which is generally fixed and assumed to have uninterrupted power supply or accessible for maintenance. The main responsibility of base station is to collect information from various sensor nodes and process it for further dissemination/actions [2]. The sensor nodes are highly resource constrained. Energy is one of the major issues of the sensor nodes. In wireless sensor network most of the energy is consumed during transmission and it is further increased with the distance, as energy consumption is directly proportional to the square of the distance among the nodes [3]. In order to minimize energy consumption and increase lifetime of the network, distance must also be kept under consideration [4] [5]. Scalability being another issue of concern as a sensor consists of hundreds or thousands of nodes [6]. In sensor networks these issues are addressed in cluster based architecture particularly in LEACH [7]. But Leach supports only static cluster formation where cluster head runs out of energy more quickly due to extra responsibility of data aggregation. In this paper we focus on key concern areas with sensor networks like minimizing energy conservation, maximizing network lifetime, real time communication.

## II. LITERATURE SURVEY

The literature [7] LEACH (Low-Energy Adaptive Clustering Hierarchy) proposed a clustering-based routing protocol that minimizes global energy usage by distributing the load to all the nodes at different points in time. A fixed clustering based approach to prolong the lifetime of sensor networks employing data aggregation is described in literature [4]. The literature [8] RDAG is an approach of Data aggregation by reducing the number of transmissions which is an effective approach to save energy by using the concept of LEACH. The literature [5] and [9] says that energy consumption in a WSN varies with the transmission range. Reducing the transmission range will reduce the power consumed in transmitting a packet toward the neighbors.

The major drawback of LEACH [7] is that it does not support dynamic clustering so to overcome this we propose dynamic clustering protocol that refreshes clusters periodically. Cluster formation is based on residual energy, cost. Our proposed method supports sleep and wait technology where in only energy is consumed only on demand. On demand energy consumption helps in proper network utilization as well as enhancing network lifetime by minimizing energy consumption.

To best of our knowledge none of the previous work has included all these factors in a single work.

III. METHODOLOGY

A. *Proposed Dynamic Clustering Protocol (DCP)*

The proposed DCP proposes formation of clusters depending upon the respective energy level of each node. It introduces the concept of assigning different energy levels to different nodes to balance the responsibility among the nodes with in a cluster. The node with the highest energy level looks for nodes within its transmission range forms a cluster and appoints itself as the cluster head of the cluster formed .Once the cluster head is identified for a cluster, transmission of data takes place from all the other nodes to the cluster head. Cluster head behaves as the data aggregating node for that particular time interval. As soon as nodes forward the data to the cluster head they move to the wait state and remain in the sleep mode until they have something more to transfer. The proposed protocol helps in conserving energy by only allowing cluster head to communicate with other cluster heads. All other nodes except cluster head are in sleep wait so their energy is preserved. Indirectly as energy is preserved the lifetime of node is increased because lifetime of a node is defined as the time period till it is capable of transmitting data. The data when aggregated [8] at cluster head of each cluster is forwarded to the base station and the energy level of the cluster head is decremented. After a fixed interval of time the energy level of each node in a cluster is revaluated and compared with other nodes and the node having the highest energy is assigned to be the new cluster head of the cluster. This enables cluster formation even when energy and position of nodes is changing i.e. dynamic clustering. This leads to an effective utilization of energy of each node in the network. Only the nodes with highest energy levels are used for transmission and the energy of all the other nodes is conserved for future use.

B. *Cluster Formation*

The formation of clusters is based upon the respective energy level of nodes. During cluster formation several parameters such as distance to base station, cluster distance, dissipated energy are taken in simultaneously, none of the previous work encounters them in a single work.

Distance to Base Station ($D_B$) is the summation of all distances from sensor nodes to the Base station (BS). This distance is defined as follows:

$$D_B = X_{1s} + X_{2s} + X_{3s} + X_{4s} \ldots\ldots\ldots + X_{ms} \quad (1)$$

$$D_B = \sum_{i=0}^{m}(X_{is}) \quad (2)$$

Where $X_{is}$ is the sum of distance from the node $i$ to the Base Station. For a larger network, try to keep this distance minimum because most of the energy will be wasted. However, for a smaller network the nodes near to base station directly send the information may be an acceptable option.

Cluster Distance ($D_C$) is the summation of the distances from the member nodes to the cluster head and the distance from the cluster head to the BS. For a cluster with k member nodes, the cluster distance $D_C$ is defined as follows:

$$D_C = X_{1h} + X_{2h} + X_{3h} + X_{4h} + \ldots\ldots + X_{kh} \quad (3)$$

$$D_C = \sum_{i=1}^{k}(X_{ih}) \quad (4)$$

Where $X_{ih}$ is the distance from node $i$ to the cluster head. For a cluster that has large number of spreads nodes, the distance among the nodes $i$ to cluster head will be more and the energy consumption will be higher. So keep the cluster size small to reduce energy dissipation and $D_C$ should not be too large. This metric will keep control the size of the clusters.

The equation (5) shows the total distances $T_{dist}$ from cluster member $i$ to cluster head and from cluster head to BS.

$$T_{dist} = \sum_{i=1}^{m}(x_{ih} + \cdots\ldots\ldots + x_{ihb}) \quad (5)$$

The total dissipated energy "$E_{total}$" is the total energy dissipated to transfer the aggregated messages from the cluster to the BS. For a cluster with k member nodes, the total dissipated energy can be calculated follows:

$$E_{total} = \sum_{i=1}^{m}(E_{Tich} + K \times E_R + E_{Tchb}) \quad (6)$$

The first part of equation (6) shows the energy consumed to transmit messages from member nodes to the cluster head. The second part shows the energy consumed to transmit aggregated messages by cluster head to BS.

The salient features of DCP include:

- Dynamic clustering
- Highest energy node selected as cluster head
- Prolonged Lifetime
- Sleep and wait
- Energy balancing
- Time to time refresh of clusters

IV. ALGORITHM

**initialise():-** Parameters of the node are assigned values here.

**cal_distance():-** Calculates distance of each node from all other nodes and stores them in the **distance** matrix.

**energy_max():-** Determines the node with max energy from the nodes which are not a part any cluster yet and returns the **id** of that node.

**status_update():-** Marks the status of the node as active or inactive.

```
Make_cluster():

1    Loop1: For i=1 to No_of_Nodes
2       Do
3          Emax_id = energy_max()
4          if(emax_id!=-1) then
5              subsink[i]=node[emax_id].id
6              node[emax_id].ss_flag=1
7              increment cluster_id
8              node[emax_id].cluster_no= cluster_id
9              loop2: For j=1 to No_of_Nodes
10                Do
11                   if(node[j].cluster_no==0) then
12                      if(distance[emax_id][j]<=range) then
13                         node[j].cluster_no=cluster_id
14             end loop2
15   end loop1
```

In this algorithm the node with the highest energy level looks for nodes within its transmission range to form a cluster and appoint itself as the cluster head of the cluster formed. The process goes on till all the nodes are not covered under some or the other cluster.

```
DCP():
1     while true,
2        do
3           while(subsink[n]!=0)
4             do
5              if(energy_val[subsink[n]-1]<=0) then
6                 print NETWORK DEAD
7                    return false
8                 decrement energy_val[subsink[n]-1] by 10 units
9             increment n
10          end while
11   loop1: for time=1 to REFRESH_TIME
12   do
13            wait for 1 time unit
14            status_update();
15            loop2: for i=1 to No_of_Nodes
16              do
17                 if(node[i].energy<=0) then
18   print NETWORK DEAD
19   return false
20              if(node[i].ss_flag!=1) then
21               if(node[i].active==1) then
22                decrement energy_val[i] by 2 units
23                     active_status[i]=0;
24               node[i].active=0;
25                else
27                   decrement energy_val[i] by 1 unit
28                end loop2
29           end loop1
30         cluster_id=0;
31          initialise()
32         cal_distance()
33         make_cluster()
34         redisplay output
35   end while
```

In this algorithm transmission of data takes place from all the other nodes to the cluster head. Cluster head behaves as the data aggregating node for that particular time interval. As soon as nodes forward the data to the cluster head they move to the wait state and remain in the sleep mode until they have something more to transfer. After some interval of time called refresh time clusters are reformed and the process repeats.

## V. WORKING SCENARIO

Cluster head nodes: filled circles ●
Active nodes: dark empty circle ◉
Inactive nodes: light empty circle ○

The different colours represent the different clusters formed.

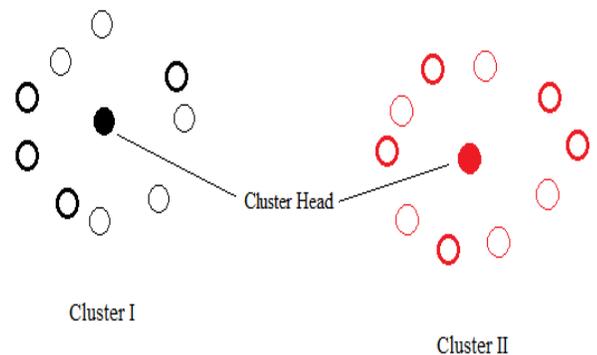

Fig.1. Cluster Representation

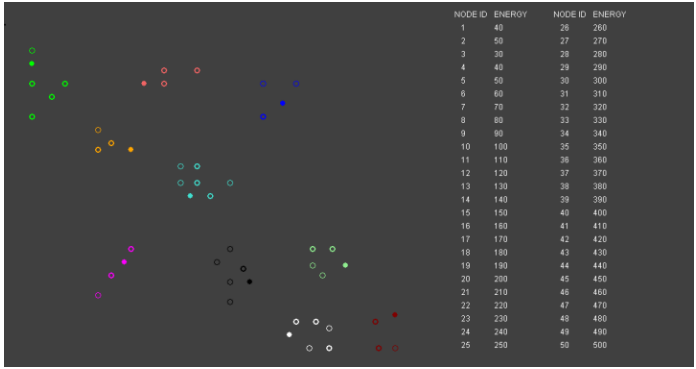

Fig.2.1. Initial clusters *(Active stage)*

Fig. 2.1 represents that clusters are formed on the basis of energy levels and the node with the highest energy level looks for nodes within its transmission range and forms a cluster and appoints itself as the cluster head of the cluster formed. The filled circle in every cluster represents the cluster head of that cluster and dark empty circle represents the active node and light empty circles are the inactive nodes. The energy values of all the nodes are also displayed along with the nodes ids.

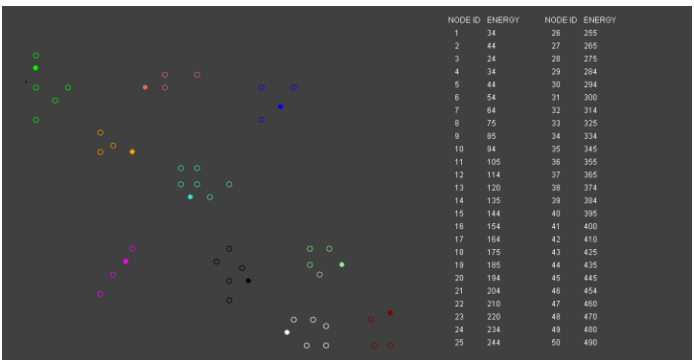

Fig.2.2. Inactive stage

Fig. 2.2 represents the status of the clusters and nodes after nodes have finished transmission of data and all the nodes are in inactive state. The energy values of all the nodes are also displayed along with the nodes ids.

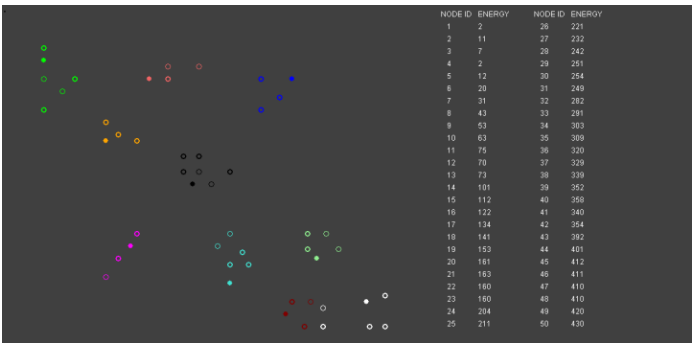

Fig.2.3. New clusters after refresh time *(Active stage)*

Fig. 2.3 represents the nodes after refreshing time when new clusters are formed and new cluster heads are appointed. The new decremented energy values of the nodes are displayed.

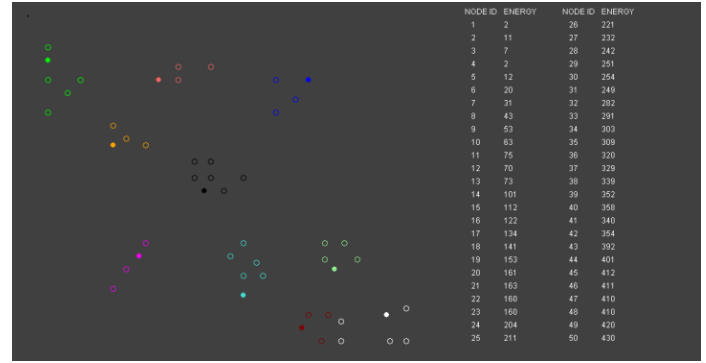

Fig.2.4. Inactive stage

Fig. 2.4 represents the status of the clusters and nodes after nodes have finished transmission of data and all the nodes are in inactive state. The energy values of all the nodes are also displayed along with the nodes ids.

## VI. SIMULATION RESULTS

We simulated and evaluated the performance of the proposed protocol with the existing Leach. All the simulations are done in MATLAB in order to validate the efficiency of proposed protocol.

| No. | Parameter | Value |
|---|---|---|
| 1. | Simulation Area | 1000x1000 |
| 2. | No. of nodes | 450 |
| 3. | Radio Propagation Model | Two way ground |
| 4. | Channel Type | Wireless Channel |
| 5. | Antenna Model | Antenna/Omniantenna |
| 6. | Energy Model | Battery |
| 7. | Round Duration Time | 10s |
| 8. | Initial Energy of each node | 0.5J |

TABLE1. SIMULATION PARAMETERS

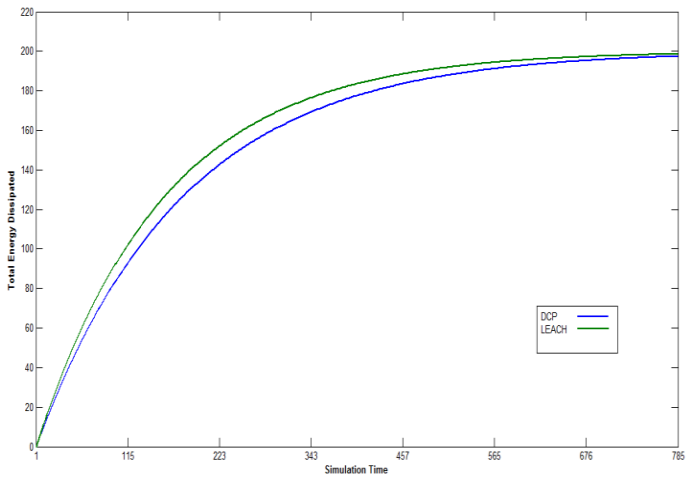

Fig.3.1. Total energy dissipated in LEACH and our scheme *(DCP)*

Fig. 3.1 shows the total energy consumption of the network over simulation time. The simulations prove DCP to be better than Leach in terms of energy consumption. At the start of simulation performance of Leach and DCP is comparable but as the simulation time proceeds DCP conserves a lot of energy due to its sleep and wait technology. As only the active nodes consume energy and nodes after forwarding data to cluster head go to sleep mode.

Fig. 3.2 shows the proposed method compared with the traditional Leach protocol in terms of network lifetime. As the energy consumption of DCP is high as compared to Leach and energy consumption is directly proportional to network lifetime. As node is alive till it runs out of energy so if energy is conserved its lifetime will increase which indirectly increases network lifetime

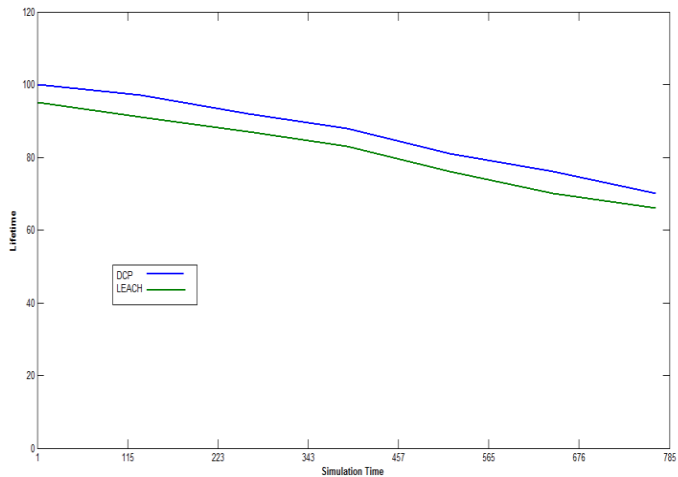

Fig.3.2. Network lifetime for LEACH and our scheme *(DCP)*

.

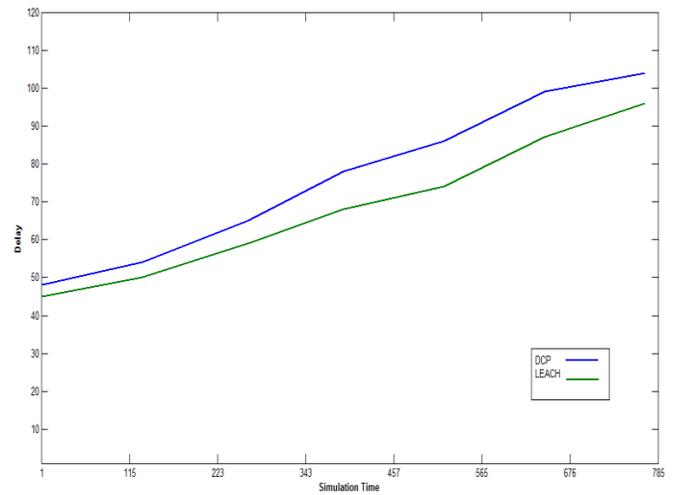

Fig.3.3. Delay in LEACH and our scheme *(DCP)*

Fig. 3.3 shows the proposed method compared with the traditional Leach protocol has a longer delay than Leach protocol primarily because DCP supports dynamic clusters and clusters are refreshed periodically. Selection of cluster head and refreshment of cluster increases delay but delay is kept proportional to real time delay (supporting real time communication) same is shown by simulation results.

## VII. CONCLUSIONS

Clustering is a useful topology-management approach to reduce the energy consumption and exploit data aggregation in wireless sensor networks. In this paper we have focused and proposed dynamic clustering where clusters are refreshed periodically and cluster head is selected accordingly. The existing protocols LEACH and RDAG support static clustering and cluster head is fixed in the entire scenario. Cluster head being used as the data aggregator node runs out of energy. So in this paper we have proposed a dynamic clustering protocol (DCP) that supports dynamic clustering with support of sleep and wait technology, where the node that needs to transmit the message is only in wake state after forwarding the message the node changes the state to sleep. By this a lot of energy is conserved enhancing network lifetime indirectly. We have simulated and compared the results in MAT lab of our proposed protocol DCP with the existing LEACH. Simulation results our proposed protocol are better in terms of energy conservation and enhancement of network life time as sleep and wait technology enhances the lifetime of nodes. Delay is increased in our DCP but still it is in real time communication limits so increased delay is not a harmful factor. In future we will compare the proposed DCP on more existing parameters. As well as we will make DCP work in real time parameters to support real time communication.

## VIII. REFRENCES